\documentclass[rsi, amsmath,amssymb, reprint]{revtex4-1}

\usepackage{graphicx}
\usepackage{dcolumn}
\usepackage{bm}
\usepackage{threeparttable}
\usepackage{multirow}

\usepackage[utf8]{inputenc}
\usepackage[T1]{fontenc}
\usepackage{mathptmx}
\usepackage{color}
\usepackage{diagbox}

\begin{document}


\title{Ultrafast time- and angle-resolved photoemission spectroscopy with widely tunable probe photon energy of 5.3 - 7.0 eV for investigating dynamics of three-dimensional materials}

\author{Changhua Bao}
\affiliation{State Key Laboratory of Low-Dimensional Quantum Physics and Department of Physics, Tsinghua University, Beijing 100084, P. R. China}

\author{Haoyuan Zhong}
\affiliation{State Key Laboratory of Low-Dimensional Quantum Physics and Department of Physics, Tsinghua University, Beijing 100084, P. R. China}

\author{Shaohua Zhou}
\affiliation{State Key Laboratory of Low-Dimensional Quantum Physics and Department of Physics, Tsinghua University, Beijing 100084, P. R. China}

\author{Runfa Feng}
\affiliation{State Key Laboratory of Low-Dimensional Quantum Physics and Department of Physics, Tsinghua University, Beijing 100084, P. R. China}

\author{Yuan Wang}
\affiliation{State Key Laboratory of Low-Dimensional Quantum Physics and Department of Physics, Tsinghua University, Beijing 100084, P. R. China}

\author{Shuyun Zhou}
 \email{syzhou@mail.tsinghua.edu.cn}
 \affiliation{State Key Laboratory of Low-Dimensional Quantum Physics and Department of Physics, Tsinghua University, Beijing 100084, P. R. China}
\affiliation{Frontier Science Center for Quantum Information, Beijing 100084, P. R. China}

\date{\today}

\begin{abstract}
Time- and angle-resolved photoemission spectroscopy (TrARPES) is a powerful technique for capturing the ultrafast dynamics of charge carriers and revealing photo-induced phase transitions in quantum materials.  However, the lack of widely tunable probe photon energy, which is critical for accessing the dispersions at different out-of-plane momentum $k_z$ in TrARPES measurements,  has hindered the ultrafast dynamics investigation of 3D quantum materials such as Dirac or Weyl semimetals. Here we report the development of a TrARPES system with a highly tunable probe photon energy from 5.3 to 7.0 eV. The tunable probe photon energy is generated by the fourth harmonic generation of a tunable wavelength femtosecond laser source by combining a $\beta$-BaB$_2$O$_4$ (BBO) crystal and a KBe$_2$BO$_3$F$_2$ (KBBF) crystal. High energy resolution of 29 - 48 meV and time resolution of 280 - 320 fs are demonstrated on 3D topological materials ZrTe$_5$ and Sb$_2$Te$_3$. Our work opens up new opportunities for exploring ultrafast dynamics in 3D quantum materials.
\end{abstract}

\maketitle

\section{\label{sec:level1}INTRODUCTION}

Revealing the ultrafast dynamics of quantum materials can provide critical information into the non-equilibrium state and possible light-induced emerging phenomena \cite{Zhou2021Rev,Sentef2021}. Time- and angle-resolved photoemission spectroscopy (TrARPES) \cite{Lanzara2016,ShenRMP2021} has been a powerful technique for capturing the ultrafast carrier dynamics and photo-induced phase transitions with energy-, momentum- and time-resolution. In the past decades, major progress has been made in TrARPES instrumentation, including extending the pump wavelength to mid-infrared \cite{Gedik2013} or terahertz \cite{Huber2018}, improving the energy resolution or time resolution \cite{Perfetti2012,Shin2014RSI,BauerRSI2016,ZX2020,Zhou2021RSI} etc.  Developments in the probe source using high harmonic generation (HHG) \cite{Bauer2007,KanidlNC2015,GedikHHGNC2019}, or free electron laser \cite{FEL2020} with a higher probe photon energy have also been achieved.   While the power of TrARPES has been demonstrated in various quasi-two-dimensional materials such as high temperature superconductors, topological insulators and graphene \cite{Lanzara2016, ShenRMP2021,Zhou2021Rev} to extract the ultrafast dynamics, so far, TrARPES studies of 3D materials such as 3D Dirac or Weyl semimetals,  where intriguing light-induced dynamics and phase transitions have been predicted \cite{Wang2016,Lee2016,Rubio2017,weber2021Rev},
have been restricted.  The challenge for TrARPES measurements of these 3D materials \cite{DingHRMP2021} is that, the Dirac or Weyl nodes exist only in isolated  out-of-plane momentum ($k_z$) points, and a highly tunable photon energy is required to access the dispersions of interests at different $k_z$ with a high precision.

The probe beam in TrARPES systems is typically generated by fourth harmonic generation (FHG) using nonlinear optical crystals $\beta$-BaB$_2$O$_4$ (BBO) with a cut-off photon energy of 6.3 eV \cite{Perfetti2012}.  Nonlinear optical crystal KBe$_2$BO$_3$F$_2$ (KBBF) crystal with prism-coupled technique (KBBF-PCT) can be used to generate high-brightness table-top vacuum ultraviolet (VUV) light source \cite{Chen1996_1,Chen1996_2,chen2002,ChenCT2008,Kolis2008,Chen2009Rev} up to 7.0 eV with excellent performance for ARPES measurements \cite{Shin2008,ZhouXJ2008,Kaminski2014,Zhou2018Rev,xu2019rev}.
A continuously tunable probe photon energy of 5.9 - 7.0 eV has been demonstrated  in static ARPES measurements \cite{Kaminski2014}. KBBF-based TrARPES measurements have been performed at a fixed photon energy with a time resolution of $\sim$ 1 ps \cite{Zhang2019}, and the time resolution has been improved to 400 - 600 fs in our recent works \cite{ZhouMBT2021,Zhou2021KekuleTr}.   Implementing TrARPES with a highly tunable probe wavelength while achieving a high time resolution is critical for further extending the ultrafast dynamics studies to 3D materials.

In this work, we report the development of a TrARPES system with a continuously tunable probe photon energy by combining KBBF-PCT based FHG and a tunable wavelength femtosecond laser source. We achieve a highly tunable probe photon energy from 5.3 to 7.0 eV, with energy  resolution of 29 - 48 meV and time resolution 280 - 320 fs. The high performance of the TrARPES system is demonstrated on 3D topological materials ZrTe$_5$ and Sb$_2$Te$_3$.

\section{EXPERIMENTAL SETUP}

\begin{figure*}
	\includegraphics[]{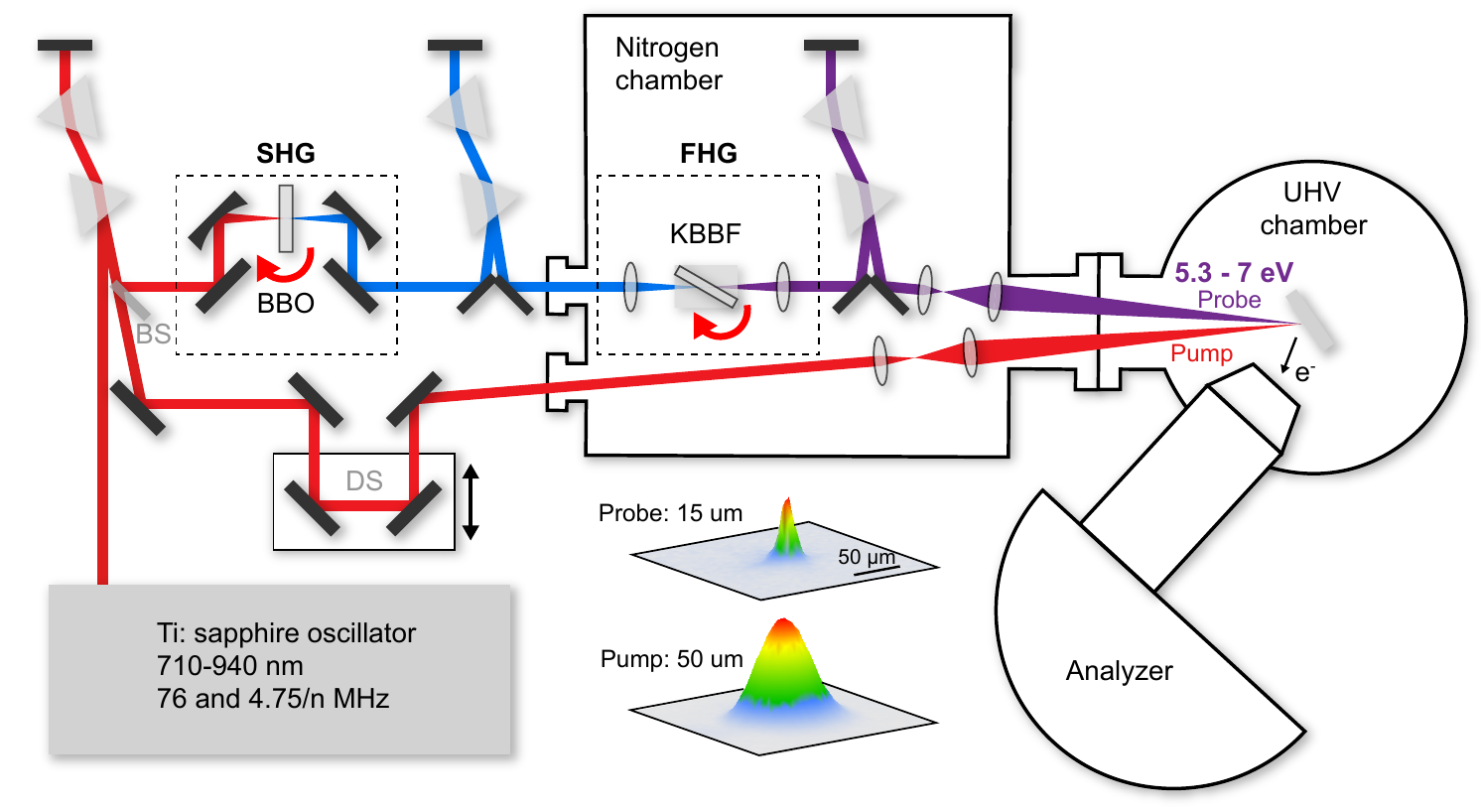}
	\caption{\label{fig:epsart} A schematic layout of TrARPES system with tunable vacuum ultraviolet laser. BS: beam splitter, DS: delay stage, UHV: ultra-high vacuum. Transverse intensity profiles of pump and probe beams. The beam sizes are defined by full width at half maximum (FWHM).}
\end{figure*}

Figure 1 shows an overview of the TrARPES system. The fundamental beam (FB) with wavelength of 710 - 940 nm is generated by a Ti:sapphire oscillator (Coherent Mira HP) with an output power of 2 - 4 W and a repetition rate of 76 MHz. The repetition rate can be reduced to 4.75/n (n is an integer) MHz by an external pulse picker.  Such flexibility allows to choose an optimum repetition rate with high data acquisition efficiency while avoiding sample damage problem induced by average heating. The laser beam is firstly compressed by a prism pair compressor to compensate for the chirp from the laser cavity, and an optimum pulse duration of 50 - 80 fs is achieved for the entire wavelength range. A beam splitter splits the FB into two branches: 50\% for the pump beam, and the other 50\% is used to generate the probe beam through FHG. Compared to an amplifier which gives a larger pulse energy, here a higher fraction of the FB is needed to generate the high-brightness probe beam for high measurement efficiency.

The pump beam goes through a delay stage which is used to change the delay time between the pump and probe pulses. Several silver mirrors are used to steer the beam onto the sample in the ultra-high vacuum (UHV) chamber for photoexcitation. Before entering the UHV chamber, the pump beam is focused by a lens pair with focal length of f = 50 and 200 mm to obtain a beam size of 50 $\mu$m as shown in Fig.~1. A maximum pump fluence of 1 $\rm mJ/cm^2$ can be applied on the sample surface.

The other 50\% of the FB is used to generate the probe beam. It is first focused onto a second harmonic generation (SHG) BBO crystal with type I phase matching ($\theta= 29.2^\circ$, thickness 0.5 mm) by an off-axis parabolic mirror (aluminum coating, f = 50 mm) as shown in Fig,~2(a).  The choice of an off-axis parabolic mirror instead of a lens is to reduce the possible chirp. The output laser beam is then collimated by another off-axis parabolic mirror and sent to a prism pair compressor to achieve optimum pulse duration for the second harmonic (SH) beam at the KBBF crystal (Fig.~1). We would like to point out that in our previous setup \cite{Zhou2021RSI} where the FHG is achieved by passing the SH through the BBO crystal directly without generating significant chirp, compression of SH is not so critical, while in the current setup, the SH passes through a large-size prism made of fused silica with a length of 15 mm before the KBBF crystal, and the chirp is significant. For example, a 50 fs SH pulse at 400 nm is elongated to 95 fs after passing through the prism. Therefore, pre-compensation for the chirp is required to achieve the optimal pulse duration for the fourth harmonic (FH) laser. The fundamental and second harmonics lasers are separated by the first prism in the compressor (see the separated beam spots in Fig.~2(a)), where the residual fundamental laser beam is dumped.

The SH laser beam is focused by a lens (ultra-violet fused silica, f = 50 mm) onto the KBBF-PCT device \cite{ChenCT2008} with type I phase matching (thickness of KBBF is 1 mm) to generate the probe beam as shown in Fig.~2(b). The SH and FH laser beams are separated by the prism
behind the KBBF crystal as shown in Fig.~2(b), to dump the residual SH laser beam.
 The FH laser beam is subsequently collimated by a CaF$\bm_2$ lens with f = 100 mm to ensure the long-distance propagation. Then it goes into a prism pair compressor for pulse duration compression and is finally focused by a lens pair (CaF$\bm_2$, f = 50 and 150 mm) to obtain a probe beam size of 15 um on the sample surface. All the FH related components are placed inside a nitrogen purged chamber to avoid absorption of VUV laser  for photon energy above 6.7 eV by the oxygen \cite{Murray1953}, water vapor, carbon oxide and organic molecular in the air. The photon energies of the pump and probe beams are related by $h\nu_{probe}=4h\nu_{pump}$ and cannot be tuned independently. However, the change in the pump photon energy is much smaller (from 1.33 eV to 1.75 eV), and the effect on the dynamics is negligible unless there is any resonant pumping. 

The sample is mounted on a motorized manipulator in an ultrahigh vacuum (UHV) chamber with a vacuum better than $5\times10^{-11}$ Torr. The manipulator has six degrees of freedom, allowing full control of the sample position and orientation. The sample is electrically insulated from the ground and connected to a picoammeter (Keithley 6485) to monitor the photocurrent. The energy and momentum of photoelectrons are measured by an electron analyzer (Scienta, DA30-L-8000).

\section{Tunable probe photon energy with wide tunability}

\begin{figure*}
	\includegraphics[]{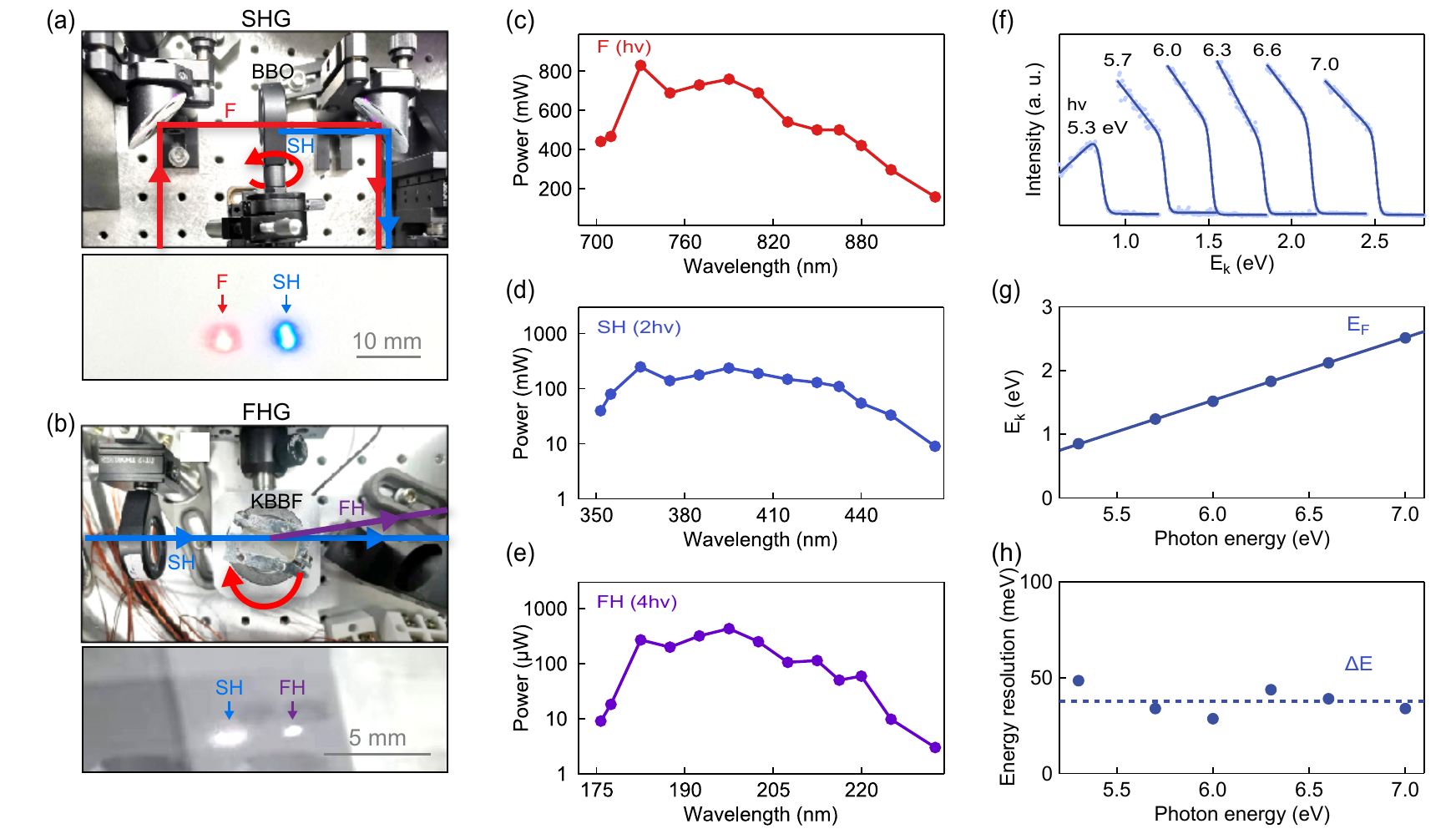}
	\caption{Probe photon flux and energy resolution. (a) Photos of SHG setup and beam spots on a white paper. (b) Photos of FHG setup and beam spots on a glass. (c) Fundamental laser power for SHG input as a function of fundamental wavelength. (d) Generated SH laser power as a function of SH wavelength. (e) Generated FH laser power as a function of FH wavelength. (f) Integrated EDCs on polycrystalline copper using different photon energies at 80 K. (g) Extracted kinetic energy of Fermi energy as a function of photon energy with a linear fitting. (h) Extracted energy resolution at different photon energies. The dashed line indicates the average level.}
\end{figure*}

The tunable probe photon energy is realized by adjusting the wavelength of the FB and corresponding phase matching angle of nonlinear BBO and KBBF crystals. To make sure the phase matching in the entire wavelength range, the BBO is installed on a large-range rotation stage to allow rotation as indicated by the red circular arrow in Fig.~2(a), and the KBBF-PCT device is installed on a piezoelectric driven large-range rotation stage as indicated by the red arrow in Fig.~2(b). The input FB beam before the SHG crystal is 160 - 800 mW at wavelength of 710 - 940 nm as shown in Fig.~2(c) and the output power of SH laser is 9 - 250 mW at wavelength of 355 - 470 nm as shown in Fig.~2(d).  The corresponding SH efficiency is 6 - 30\%.
 The power of final FH laser is 3 - 430 $\rm\mu W$, which corresponds to a photon flux of 4$\times$10$^{12}$ - 4$\times$10$^{14}$ $\rm s^{-1}$ at photon energy of 5.3 - 7.0 eV (235 - 177.5 nm) as shown in Fig.~2(e), which is still much higher than typical HHG light source (<10$^{11}$$\rm s^{-1}$) \cite{Ernstorfer2019}. Therefore, a high-brightness probe laser beam with tunable photon energy of 5.3 - 7.0 eV is achieved for TrARPES measurements. We note that for typical BBO-based TrARPES, the highest photon energy is 6.05 eV in  the 2$\omega$ + 2$\omega$ scheme or 6.3 eV in the $\omega$ + 3$\omega$ scheme \cite{Perfetti2012}. By using KBBF-PCT, we can extend it to cover the previously inaccessible  probe photon energy range of 6.3 - 7.0 eV.
In addition to accessing a larger range of $k_z$ momentum, the higher photon energy also allows to probe a larger energy range below the Fermi energy.

The energy resolution at different probe photon energies is measured on a polycrystalline copper at 80 K. The angle-integrated energy distribution curves (EDCs) for photon energies from 5.3 to 7.0 eV are shown in Fig.~2(f). The Fermi energy ($E_F$) and energy resolution ($\Delta E$) can be extracted by fitting EDCs with Fermi-Dirac function: $f(E)=1 /(1+e^{\frac {E-E_F}{\Delta_{wid}/4}})$, where $\Delta_{wid}$ corresponds to width of the Fermi edge and is related to $\Delta E$ by $\Delta_{wid}=\sqrt{\Delta_E^2+\Delta_{T}^2}$.  Here $\Delta_{T}$ = 4$k_BT$ = 28 meV at the measurement temperature of T = 80 K is induced by the thermal broadening (defined by 12\% to 88\% of the edge), which could be reduced when cooling down to a lower temperature, and $\Delta E$ is the energy resolution with contributions from the probe source, analyzer and sample broadening. 
The extracted  $\Delta_{wid}$ from the fitting is 40 - 56 meV, and the energy resolution after subtracting the contribution from the thermal broadening $\Delta_{T}$ is 29 - 48 meV for the entire photon energy range as shown in Fig.~2(h). The energy resolution is limited by the time resolution through the uncertainty principle and could be improved by sacrificing the time resolution, for example, a higher energy resolution of 16 meV \cite{Zhou2021KekuleTr} can be achieved by compromising the time resolution.

\begin{figure*}
	\includegraphics[width=18cm]{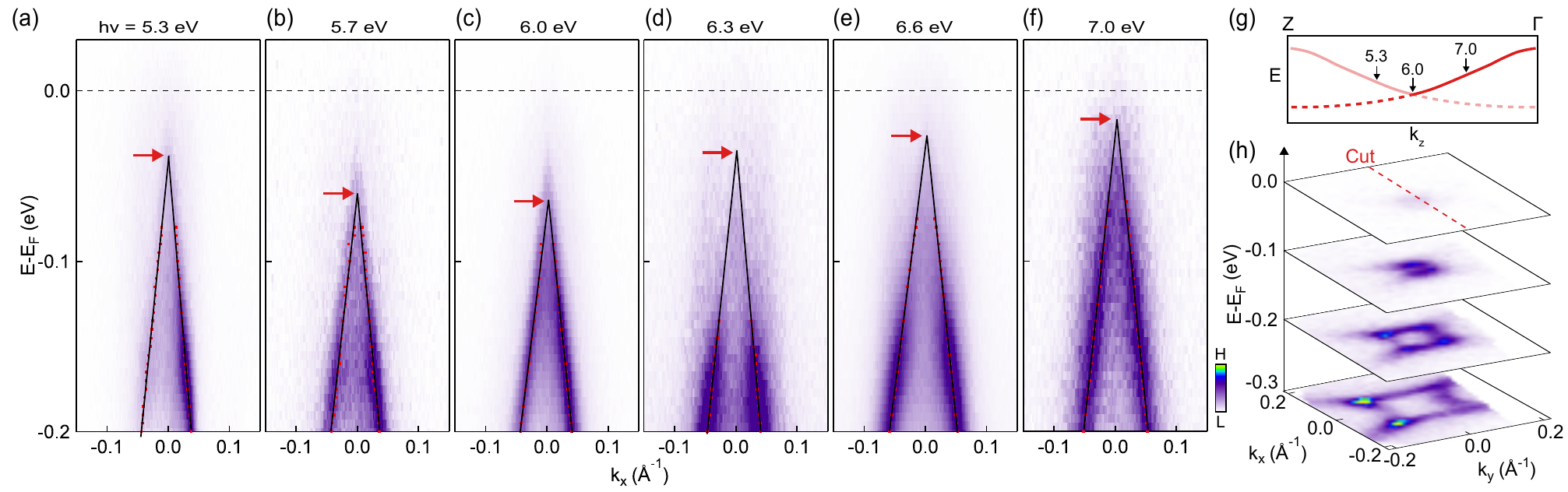}
	\caption{The demonstration of tunable photon energy on ZrTe$_5$ single crystal.  (a)-(f) Dispersion images along the direction as indicated in (h) using different photon energies. The red dots are extracted from momentum distribution curves peaks, and the black lines are the extrapolated dispersion. The red arrows indicate the top of the valence band. (g) The schematic band dispersion along k$\rm _z$ direction. The red curve is original dispersion and the light red curve is folded dispersion. The black arrows indicate corresponding k$\rm _z$ for photon energies of 5.3, 6.0 and 7.0 eV. (h) Intensity maps at different energies using photon energy of 7.0 eV. The measurement temperature is 80 K.}
\end{figure*}

The suitability of such highly tunable probe photon energy source for  3D quantum materials is demonstrated on ZrTe$_5$, which is a topological material \cite{Zhong2014,Vala2016} with strong  k$_z$ dispersion \cite{Grioni2016,Crepaldi2016,ZX2017,XJ2017}. Figure 3(a)-(f) shows dispersion images measured along the $\Gamma$ - X direction (indicated by the dashed line in Fig.~3(h)) at a few selected photon energies from 5.3 to 7.0 eV. This photon energy range corresponds to k$_z$ = 0.70 - 0.87 $c^*$ \cite{Grioni2016}, covering 34$\%$ of $\Gamma$ - Z (half of the k$_z$ Brillouin zone) as shown in Fig.~3(g).  For typical 3D topological materials, the lattice constant is 7.25 \AA ~ for ZrTe$_5$ \cite{Grioni2016}, 9.7 \AA ~for Na$_3$Bi \cite{Chen2014}, 12.6 \AA ~for Cd$_3$As$_2$ \cite{Cava2014} and 11.6 \AA ~for TaAs \cite{Ding2015}, so the tunable photon energy of 5.3 -7.0 eV can cover around 34$\%$-60$\%$ of the half of the k$_z$ Brillouin zone for typical 3D topological materials (assuming an inner potential of 7.5 eV \cite{Grioni2016}). 

A significant change is clearly identified in the top of the valence band as indicated by red arrows in Fig.~3(a)-(f), showing strong $k_z$ dispersion.  The shift of the top of the valence band toward the Fermi energy from photon energy of 6.0 to 7.0 eV is consistent with previous reports \cite{XJ2017} due to the strong k$\rm _z$ dispersion along the $\Gamma$ - Z direction as illustrated schematically in Fig.~3(g). From 6.0 to 5.3 eV, the top of the valence band moves up, possibly due to the 1/2 reconstruction induced band folding along the k$\rm _z$ direction \cite{Grioni2016} as indicated by the light red curve in Fig.~3(g).  Figure 3(h) shows intensity maps measured at different energies using a probe photon energy of 7.0 eV, where an expanding rectangular pocket is clearly identified, in agreement with previous reports \cite{ZX2017,XJ2017}.

\section{Ultrafast dynamics and time resolution of TrARPES}

\begin{figure*}
	\includegraphics[width=18 cm]{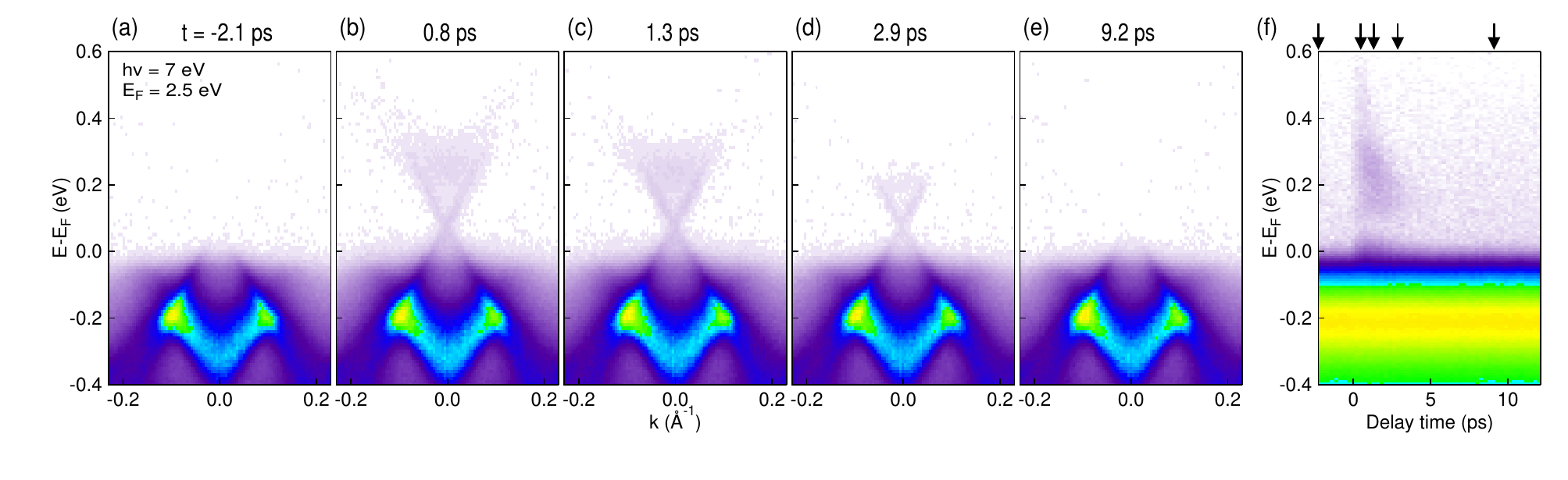}
	\caption{The demonstration of TrARPES on Sb$_2$Te$_3$ film. (a)-(e) Snapshots of electronic dispersion at different delay times as indicated by black arrows in (f) using probe photon energy of 7.0 eV (Fermi energy is 2.5 eV) and pump photon energy of 1.75 eV. (f) Corresponding integrated energy distribution curves as a function of delay time. The pump fluence is 80 $\mu$J/cm$^2$ and the temperature is 80 K. }
\end{figure*}

The capability of the developed TrARPES system for capturing the ultrafast dynamics is demonstrated on topological insulator Sb$_2$Te$_3$ at a probe photon energy of 7.0 eV. The sample is a thin film grown by molecular beam epitaxy (MBE) with a slight hole doping and only the dispersion below the Dirac point is observed at negative time delay (no pump) in Fig.~4(a). Upon photoexcitation, electrons are excited to the unoccupied states above the Fermi energy and the whole Dirac cone and bulk conduction band are observed  in Fig.~4(b). The photoexcited electrons relax to equilibrium state in several picoseconds as shown in Fig.~4(b)-(e). Furthermore, the evolution of photoexcited electrons is also measured by changing the delay time between the pump and probe laser pulses. As shown in Fig.~4(f), the electronic states at high energy are dominated by direct photoexcitations and has a much shorter lifetime, which can be used to extract the time resolution.

The time resolution is obtained by analyzing the temporal evolution of photoexcited electrons at high energy measured on Sb$_2$Te$_3$, which is dominated by the direct photoexcitation \cite{Zhou2021RSI}. Figure 5(a) shows the images measured as a function of energy and delay time. By integrating over the energy range indicated by dashed lines in Fig.~5(a), the intensity curve  in Fig.~5(b) shows a fast rising edge. Fitting the data by a Gaussian function convolved with the product of the step function and single-exponential function allows to extract the time resolution \cite{Zhou2021RSI}: $$I(t)=A(1+erf(\frac t {\Delta t}-\frac {\Delta t}{2\tau}))e^{-\frac {t} \tau})+B,$$ where $\Delta t$ is the width of the rising edge and $\tau$ is the relaxation time. The time resolution (FWHM) is $2\sqrt{\ln 2}\Delta t$ and extracted to be 280 $\pm$ 20 fs for 5.9 eV. The time resolution is also extracted to be 280 $\pm$ 10 and 320 $\pm$ 80 fs for probe photon energy of 6.4 and 7.0 eV respectively as shown in Fig.~5(c)-(f). For 7.0 eV, a lower energy range is used to extract the time resolution due to the negligible excitation for the high-energy band. This explains the slower relaxation in Fig.~5(f), and suggests that the real time resolution could be slightly better than the extracted value.

Finally, we would like to discuss the effect of the two prism compressors which are critical for achieving the time resolution discussed above.  Without such compressors, TrARPES can also be performed on KBBF-PCT based systems, yet with a somewhat compromised time resolution.  For example, in our previous work, the time resolution is 480 fs at 6.2 eV \cite{Zhou2021KekuleTr} and 610 fs at 6.66 eV \cite{ZhouMBT2021}. Here by adding the prism compressors, we can greatly improve the time resolution and achieve an optimum resolution of 280 - 320 fs. The combined advantages of highly tunable probe photon energy and high time resolution make it a fantastic system for ultrafast dynamic studies of 3D quantum materials, e.g. topological Dirac, Weyl and nodal line semimetals.


\begin{figure*}
	\includegraphics[width=18 cm]{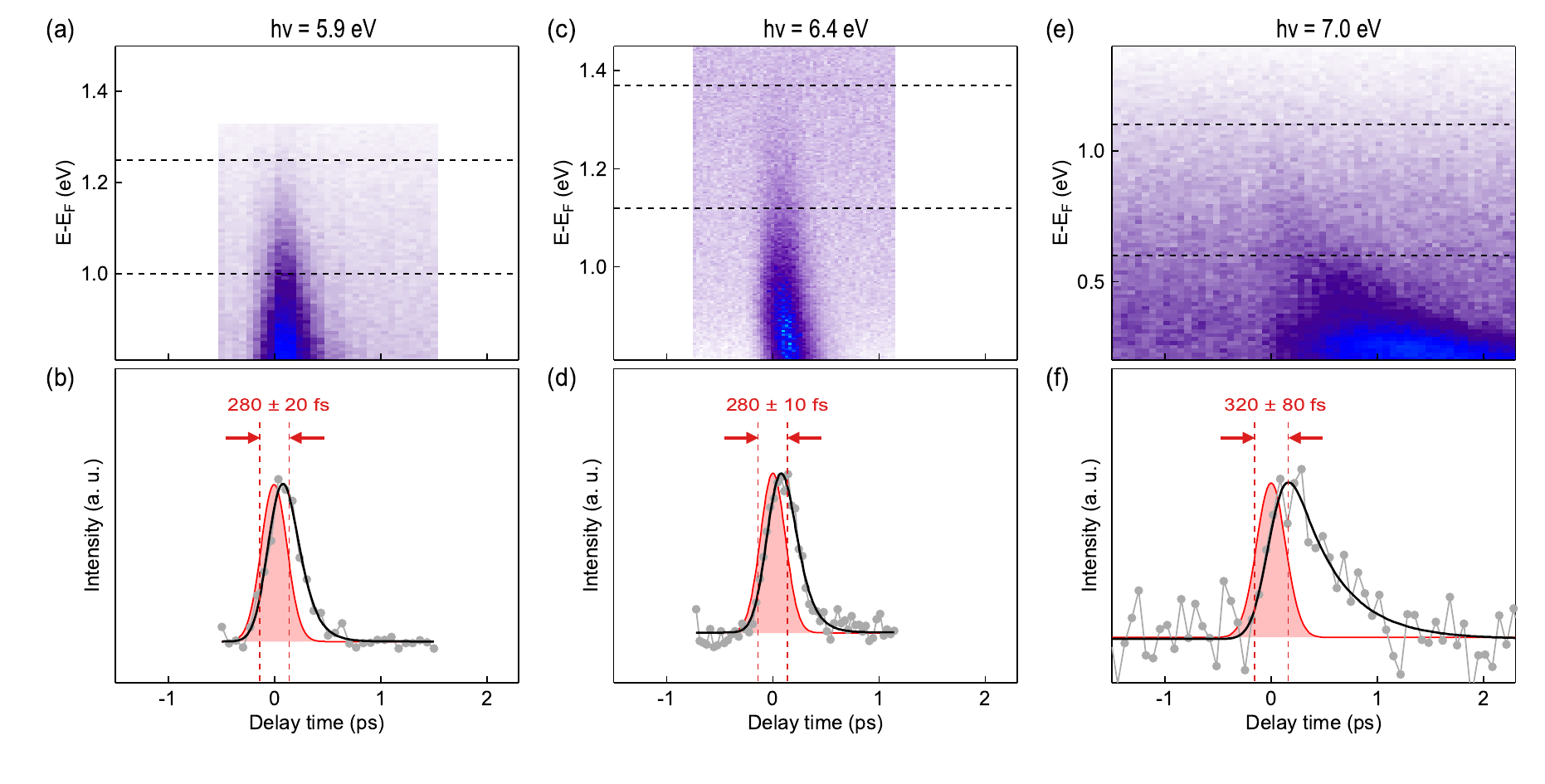}
	\caption{The time resolution of the TrARPES system. (a),(c),(e) Integrated energy distribution curves as a function of delay time at probe photon energy of 5.9 (a), 6.4 (c) and 7.0 eV (e). The pump photon energies are 1.48, 1.60 and 1.75 eV for (a),(c),(e) respectively. (b),(d),(f) Integrated intensity as a function of delay time for regions between dashed lines in (a),(c) and (e) with a single-exponential relaxation fitting. Red curves are the fitted Gaussian peaks for extracting the time resolution.}
\end{figure*}

\section{Conclusions}

In summary, we develop a KBBF-based TrARPES system with a widely tunable probe photon energy VUV laser covering 5.3 - 7.0 eV. The energy resolution is 29 - 48 meV and the time resolution is 280 - 320 fs. The TrARPES system is demonstrated on 3D topological materials. Our work opens up new opportunities for exploring ultrafast light-mater interaction in 3D quantum materials \cite{Wang2016,Rubio2017,Zhou2021Rev,Sentef2021}.

\begin{acknowledgments}
This work was supported by the National Natural Science Foundation of China (Grant No.~11427903, 11725418), the National Key R\&D Program of China (Grant No. 2016YFA0301004,  2020YFA0308800), Beijing Advanced Innovation Center for Future Chip (ICFC) and Tohoku-Tsinghua Collaborative Research Fund.
\end{acknowledgments}

\textbf{Data Availability Statement:} The data that supports the findings of this study are available within the article.

%

\end{document}